Superconductivity in PbO-type Fe chalcogenides

Yoshikazu Mizuguchi and Yoshihiko Takano

1. National Institute for Materials Science, 1-2-1 Sengen, Tsukuba, 305-0047, Japan
2. Japan Science and Technology Agency-Transformative Research-Project on Iron-Pnictides (JST-TRIP), 1-2-1 Sengen, Tsukuba, 305-0047, Japan
3. University of Tsukuba, 1-1-1 Tennodai, Tsukuba, 305-8571, Japan

Abstract

PbO-type Fe chalcogenide has been drawing much attention as the Fe-based superconductor with the most simple crystal structure. Whereas FeSe is an intrinsic superconductor, FeTe, while having a structure analogous to FeSe, exhibits antiferromagnetic ordering. The relationship between antiferromagnetism and superconductivity provides key information to understand better superconductivity in Fe chalcogenides. Furthermore the significant pressure effect on $T_c$ in the Fe-chalcogenide superconductors, which is likely to be correlated with the crystal structure and/or magnetism, is important in elucidating the mechanism of Fe-based superconductivity. Due to the simple structure and composition, Fe-chalcogenide superconductor is one candidate for applications to such areas as superconducting wires and thin films.

1. Introduction

Within the 100 years since the discovery of superconductivity in Hg [1], several types of superconductors have been discovered. Although a large number of superconductors exist, only two systems achieve superconductivity at a high transition temperature ($T_c$). The system that has the highest $T_c$ is the Cu-oxide (cuprate) superconductor, which was discovered in 1986 [2-5]. A layered crystal structure built of $CuO_2$ planes is a feature common to all superconductors belonging to this family. The second system is the Fe-based superconductor with a $T_c$ above 50 K, which was discovered in 2008 [6-8]. An Fe-based superconductor also exhibits a layered structure. Figure 1 (a), (b) and (c) are the crystal structures of LaFeAsO, $BaFe_2As_2$ and FeSe, respectively. A structure essential to Fe-based superconductivity is an Fe square lattice as depicted in Fig. 1 (d) [6-13].

In regards to superconductivity applications, e.g. the superconducting wires used in a high-field superconducting magnet, high-$T_c$ compounds are not in wide use, and the most commonly employed material is NbTi, a binary superconductor with a $T_c$ of 10 K. This can be understood by noting the desired features for producing high quality superconducting wires are the ease of fabrication process, low cost, and the abundance of materials. Binary superconductors are suitable in these respects. Therefore the discovery of a binary superconductor with high $T_c$, large critical current density ($J_c$) and high upper critical field ($H_{c2}$), will clearly enhance the potential applicability of superconductivity. At present, a major candidate material that fulfils these requirements is $MgB_2$ with a $T_c$ of 39 K [14,15]. It is worth noting that while $MgB_2$ had been a well-known commercial material for a long time, it was only in 2001 that it was finally found to be a superconductor.

Recently a new binary Fe-based superconductor, FeSe, was discovered [11]. Similar to the case of $MgB_2$, PbO-type FeSe had been in wide circulation as a commercial material [16], but its superconductivity was discovered only after the Fe-based superconductors came into focus. This was soon followed by the discovery of the superconductors $FeTe_{1-x}Se_x$ and $FeTe_{1-x}S_x$ [17-19]. Owing to its crystal structure, which is the simplest among Fe-based superconductors, Fe chalcogenides have been drawing much attention from theorists and experimentalists alike. Below we provide a summary on the structural and physical properties of the Fe chalcogenides. A topic of central interest on these materials is the large pressure effect on superconductivity. Highlighting the superconducting properties under pressure, we will focus on the relation between the structure, magnetism and superconductivity in the Fe-based superconductors. We also describe current work on thin-film and wire fabrication.

## 2. Physical properties of Fe chalcogenide

### 2-1. FeSe

FeSe has an anti-PbO-type crystal structure composed solely of $Fe_2Se_2$ layers as shown in Fig. 1(c). The PbO structure is one of the stable phases of the Fe-Se binary compounds [16]. One can therefore obtain a polycrystalline sample of PbO-type FeSe using a conventional solid state reaction method. Although the obtained sample often contains the hexagonal Fe-Se phase, annealing in an evacuated tube at 300 – 400 ºC transforms it to the tetragonal (PbO-type) phase, as shown in Fig. 2 [20].

Figure 3 shows the temperature dependence of the lattice constants of FeSe. With decreasing temperature, PbO-type FeSe exhibits a structural transition from tetragonal to orthorhombic at 70 – 90 K [21,22]. This structural transition is not accompanied by any magnetic transition. FeSe does not exhibit any magnetic ordering [20,23] and has a superconducting phase with $T_c^{onset}$ ~ 12 K and $T_c^{zero}$ ~ 8.5 K [11,21]. Superconductivity occurs in the orthorhombic phase. Figure 4 shows the temperature dependence of resistivity for $FeSe_{0.88}$, taken from the first report of superconductivity in FeSe by F. C. Hsu et al. [11]. A high $H_{c2}$, which is a common feature to all Fe-based superconductors, was reported. T. M. McQueen et al. has reported an extreme sensitivity of superconductivity in FeSe to stoichiometry, which suggested that $Fe_{1.03}Se$ does not show superconductivity whereas $Fe_{1.01}Se$ does, as shown in Fig. 5 [20].

### 2-2. FeTe

PbO-type structure is a stable phase of Fe-Te binary compounds [24]. PbO-type FeTe is composed of $Fe_2Te_2$ layers and the excess Fe that occupies the interlayer sites as depicted in Fig. 6(a). The actual composition can thus be described as $Fe_{1+d}Te$ ($d$ = 0.07 ~ 0.25) [24-26]. As is the case of FeSe, FeTe also exhibits a structural transition around 70 K; this however is accompanied by antiferromagnetic ordering [25]. Figure 7 shows the temperature dependence of resistivity for FeTe and FeSe [17,18,27]. In contrast to FeSe, FeTe does not show metallic behavior above 70 K, i.e. in the tetragonal phase. Below 70 K, a sudden drop of resistivity is observed, and metallic behavior appears. Figure 8 shows the Mössbauer spectra for FeTe [23]. A clear magnetic sextet is observed at low temperatures; the corresponding hyperfine field is $H_{hf}$ = 103.4(11) kOe at 4.2 K. The amount of excess Fe affects the structural and physical properties of FeTe [26,27]. When the concentration of excess Fe is low, the structural transition occurs between the tetragonal and the monoclinic phases, and a commensurate magnetic structure is observed (Fig. 6(c,e)). In contrast, with a higher $d$, a structural transition between the tetragonal and the orthorhombic phases occurs and an incommensurate magnetic structure is observed (Fig. 6b, d).

As mentioned above, FeTe exhibits antiferromagnetic ordering, but the magnetic spin structure of FeTe has a wave vector of $Q = (\pi, 0)$, and differs from that of FeAs-based compounds, which is given by Q = (π, π), as described in Fig. 9 [25,26]. However, Fe-based superconductivity appears in both FeAs- and FeTe-based compounds upon the suppression of long-range antiferromagnetic ordering. This implies that the magnetic structure of the parent phases is irrelevant for the appearance of superconductivity in Fe-based materials. Meanwhile, however, high-$T_c$ superconductivity in FeAs-based compounds is generally believed to be associated positively with antiferromagnetic fluctuations with Q = (π, π). Interestingly, the magnetic fluctuations with Q = (π, π) have been observed in superconducting phase of FeTe$_{1-x}$Se$_x$. This suggests that the same magnetic fluctuations as in the FeAs-based materials play a crucial role in giving rise to superconductivity in Fe chalcogenides. The authors believe that FeTe exhibits a wave vector different from FeAs-based compounds due to the presence of excess Fe at the interlayer site, which prevents it from realizing its tendency to exhibit the same ordering as in the FeAs-based compounds. The relation between superconductivity and antiferromagnetic fluctuations is discussed in section 2-4.

2-3. FeS

There are several types of Fe-S binary compounds. However, the PbO-type structure is an unstable phase in Fe-S binaries and conventional solid-state reaction methods do not result in the phase. PbO-type FeS samples can be obtained only from a chemical-processed synthesis in an aqueous surrounding [29,30]. It is known that the crystal structure of tetragonal (PbO-type) FeS easily transforms under high temperatures and/or pressures into a hexagonal structure, which is more densely packed. While there are several reports at present on the structural properties of PbO-type FeS, further detailed along the this line is necessary to gain a better understanding on the mechanism of Fe-based superconductivity.

2-4. FeTe$_{1-x}$Se$_x$

Since the end members, FeSe and FeTe, have the PbO-type structure, a mixed phase FeTe$_{1-x}$Se$_x$ can also form the PbO-type structure. Furthermore, a theoretical study has predicted that the FeTe-based compounds would exhibit a higher $T_c$ than FeSe because of the stronger antiferromagnetic fluctuations [31]. Indeed, with an optimal composition, FeTe$_{1-x}$Se$_x$ shows superconductivity at 14 K, which is the highest $T_c$ at ambient pressure among the Fe chalcogenides [17,18,28]. In these systems, superconductivity appears upon the suppression of the antiferromagnetic ordering observed in FeTe by partial substitution of Te by Se.

Figure 10 shows the temperature dependence of magnetic susceptibility for FeTe$_{1-x}$Se$_x$ single crystals. The highest $T_c$ of 14 K appears at $x \sim 0.4$. Large single crystals of FeTe$_{1-x}$Se$_x$ can be grown easily using the self-flux method or Bridgman method as shown in Fig. 11 [32] and have been submitted to various measurements. Taen et al. have reported that low-temperature annealing around 400 ºC was required to make the obtained crystals superconducting as in FeSe [33].

Figure 12 is the phase diagram for $FeTe_{1-x}Se_x$ with lower excess Fe concentration [12]. The single crystals were prepared with a nominal starting composition of $Fe_{1.00}Te_{1-x}Se_x$. With increasing Se concentration, antiferromagnetic ordering is suppressed and bulk superconductivity appears. The intrinsic nature near the boundary between superconducting and antiferromagnetic phases, at $x = 0.1 \sim 0.15$, has not been established at present. The possibility of the coexistence of magnetism and superconductivity has been also suggested experimentally [34-36]. Figure 13(a-f) shows the volume fractions of superconducting and magnetic phases detected from neutron diffraction, and Fig. 13(g) is the phase diagram established from these results [35]. As mentioned above, the long-range antiferromagnetic ordering observed in the monoclinic structure possesses a commensurate wave vector while the short-range ordering observed in the orthorhombic structure has an incommensurate wave vector. Therefore, due to the sensitivity of material properties to the excess Fe concentration it is difficult to determine whether magnetism and superconductivity coexist or not. In fact, the superconducting properties are also affected by the concentration of excess Fe. As in Fig. 14, in $Fe_{1+d}Te_{1-x}Se_x$ with a higher $d$, bulk superconductivity is not observed and instead a spin-glass-like behavior is seen [27,37]. In this respect, it is difficult to accurately describe the physical properties of the $Fe_{1+d}Te_{1-x}Se_x$ system because of the large number of contributing factors and the difficulty in controlling the stoichiometry of the samples.

Although the wave vector of magnetically ordered FeTe is $Q = (\pi, 0)$, which is different from that of the FeAs-based parent compounds, $Q = (\pi, \pi)$, as depicted in Fig. 9, magnetic fluctuations in $FeTe_{1-x}Se_x$ with a wave vector of $Q = (\pi, \pi)$ was detected from neutron diffraction [38]. Figure 15 shows the $\hbar\omega - T$ dependence of magnetic scattering at the nesting vector. The spin resonance and the associated spin gap appear with superconductivity. This suggests that the spin fluctuation is relevant to superconductivity is the same in Fe chalcogenides and FeAs-based superconductors.

2-5. $FeTe_{1-x}S_x$

Partial substitution of Te by S also suppresses antiferromagnetism in FeTe and achieves superconductivity as in the case of the Se substitution for Te [19]. Figure 16 shows the temperature dependence of resistivity for $FeTe_{1-x}S_x$ prepared by melting and quenching. The anomaly corresponding to the structural and magnetic transition is suppressed with increasing S concentration, and disappears in $FeTe_{0.8}S_{0.2}$ where zero resistivity arises. While the melted samples show sharp superconducting transitions, they also contained impurity phases. In fact, the superconducting volume fraction estimated from the magnetic measurement was less than 20 %. In contrast, the polycrystalline sample of $FeTe_{1-x}S_x$, for example, prepared with a starting nominal composition of $FeTe_{0.8}S_{0.2}$, was almost a single phase. However, it did not show bulk superconductivity in spite of the fact that the long-range antiferromagnetic ordering was suppressed. Single crystals grown using a self-flux method did not show bulk superconductivity either. To discuss the reason why bulk superconductivity was not observed in those samples, we plotted the lattice constant $c$ for the melted, polycrystalline and single crystal samples in Fig. 17. The value of $c$ for melted FeTe$_{0.8}$S$_{0.2}$ is the smallest among these samples. As reported in Ref. 39, the solubility limit of S for the Te site is less than ~0.15, due to a large difference in the ionic radius between Te$^{2-}$ (211 pm) and S$^{2-}$ (186 pm). Therefore, the conventional solid-state reaction would not allow obtaining highly-doped FeTe$_{1-x}$S$_x$. To compensate for the insufficiency of S concentration, several unique post-treatment processes, which induce bulk superconductivity, have been proposed for as-grown FeTe$_{1-x}$S$_x$, which does not show bulk superconductivity.

Figure 18(a) and (b) show the temperature dependence of resistivity and susceptibility for FeTe$_{0.8}$S$_{0.2}$ poly crystals with several air-exposure time, respectively [40]. Although the as-grown sample did not show zero resistivity, the superconducting transition became sharper with longer exposure time. After 200 days, $T_c^{zero}$ reached 7.5 K. The shielding volume fraction estimated from the susceptibility after zero-field cooling was also enhanced up to 74 %, which indicates that the superconducting volume fraction was strongly enhanced by air exposure. The evolution of superconductivity in air is likely to be related to moisture in air, because the sample kept in water at room temperature alone showed superconductivity among those kept in oxygen, nitrogen, argon, and vacuum. Furthermore, the speed of evolution of superconductivity was enhanced by immersing the sample into hot water, which suggests that the surrounding temperature is an important factor for inducing superconductivity in FeTe$_{0.8}$S$_{0.2}$. Water-induced superconductivity has been reported as well in BaFe$_2$As$_2$, which is one of the parent phases of the FeAs-based superconductor [41]. Elucidation of the origin of water-induced superconductivity will provide us with key information for discovering an Fe-based superconductor of a new type. Recently, it was reported that heating the FeTe$_{0.8}$S$_{0.2}$ sample in alcoholic drinks, particularly in red wine, was more effective for inducing superconductivity than in pure water, ethanol or water-ethanol mixture; the speed of evolution of superconductivity is about 6 times as fast as water [42]. Figure 19 shows the liquid dependence of the superconducting volume fraction for the FeTe$_{0.8}$S$_{0.2}$ sample heated in various liquids: water, ethanol, water-ethanol mixture, wine, beer, whisky, Japanese sake, and shochu. The shielding volume fraction, which roughly corresponds to the superconducting volume fraction, is clearly higher than the water-ethanol mixture.

Although superconductivity is not induced when as-grown FeTe$_{0.8}$S$_{0.2}$ is kept in oxygen gas at room temperature, annealing at 200 ºC in oxygen gas rapidly induces bulk superconductivity in FeTe$_{1-x}$S$_x$ [43,44]. Figure 20 (a) shows the temperature dependence of resistivity for as-grown and oxygen-annealed FeTe$_{0.8}$S$_{0.2}$. By oxygen annealing at 200 ºC for 2 hours, zero resistivity was observed at 8.5 K. The resistivity-temperature curve for the annealed sample has a broad hump-like structure around 60 K, which is similar to that observed in optimally doped FeTe$_{1-x}$Se$_x$ superconductors, i.e., FeTe$_{1-x}$Se$_x$ with $x$ = 0.3 ~ 0.5 [32,33]. Figure 20 (b) shows the temperature dependence of susceptibility for FeTe$_{0.8}$S$_{0.2}$ annealed in oxygen at various temperatures. The diamagnetic signal was suppressed with oxygen annealing at higher temperatures of $T$ < 200 ºC. Figure 21 summarizes how the lattice changes by oxygen annealing at various temperatures. Superconductivity was induced when the lattice was compressed by oxygen annealing. The lattice constants for optimally annealed FeTe$_{0.8}$S$_{0.2}$, which shows superconductivity, are obviously smaller than that of the melted sample. The shrinkage of lattice was observed in air-exposed FeTe$_{0.8}$S$_{0.2}$ as well, implying that the lattice shrinkage is directly related to the appearance of bulk superconductivity in this system.

Evolution of superconductivity by oxygen annealing was observed for FeTe$_{1-x}$S$_x$ single crystals grown using the self-flux method. Figure 22(a) and (b) show the temperature dependence of the normalized resistivity for as-grown and oxygen-annealed FeTe$_{1-x}$S$_x$ crystals, respectively. While S substitution for Te suppresses long-range antiferromagnetic ordering, oxygen annealing does not affect the magnetic transition temperature, which suggests that the role of oxygen is different from that of S in FeTe$_{1-x}$S$_x$. A phase diagram of oxygen-annealed FeTe$_{1-x}$S$_x$ established using the single crystals is shown in Fig. 23. FeTe$_{1-x}$S$_x$ single crystals with $x$ > 0.15 were unattainable, due to the low solubility limit of S for Te. Single crystals with $x$ > 0.15 can be obtained using alternative methods, such as the flux-method or the floating-zone method, and they show bulk superconductivity without any post treatments.

## 3. Pressure effect of Fe chalcogenide

### 3-1. Pressure effect on superconducting properties of FeAs-based compounds

Pressure study for Fe-based superconductors is important for elucidating the mechanism of superconductivity in Fe-based compounds. Soon after the discovery of superconductivity in $LaFeAsO_{1-x}F_x$, Takahashi et al. has reported that the $T_c$ of $LaFeAsO_{1-x}F_x$ reached 43 K under high pressure [45], which activated research on Fe-based superconductivity. While some FeAs-based superconductors show an increase of $T_c$ under pressure (positive pressure effect), others show a decrease of $T_c$ under pressure (negative pressure effect); for example, the $T_c$ of $NdFeAsO_{1-d}$ dramatically decreases from 54 to 16 K at 18 GPa [46]. This suggests that the change in $T_c$ is related to the change in the local crystal structure under high pressure [47,48]. Furthermore, application of pressure to the parent compound of FeAs-based superconductor, which exhibits antiferromagnetic ordering, suppresses magnetism and induces superconductivity [49-52]. We therefore expect that an application of pressure suppresses the antiferromagnetic transition in FeTe and induces superconductivity.

Due to the simple crystal structure of Fe chalcogenide, composed only of Fe-chalcogen layers, studies on both physical and structural properties for Fe chalcogenides under high pressure provide us with important information about the mechanism of superconductivity in Fe-based superconductors. Here we summarize the various pressure effects of FeSe, FeTe, $FeTe_{1-x}Se_x$ and $FeTe_{1-x}S_x$. Studies on effects of strain in both thin films and bulk samples are also summarized.

### 3-2. Pressure effect of FeSe

FeSe shows the largest pressure effect on $T_c$ among the Fe-based superconductors. The value of $T_c^{onset}$ increases from 12 K to 37 K under a pressure of 4–6 GPa [53-56]. Figure 24 shows the temperature dependence of resistivity for FeSe under high pressure, which was measured using an indenter pressure cell [56]. With increasing pressure, the normal-state resistivity decreases and the $T_c$ increases dramatically. In general, the superconducting transition becomes broader under high pressure, which can probably be attributed, for example, to the inhomogeneity of the applied pressure. However the superconducting transition of FeSe becomes sharper with an application of a pressure of 1 GPa. Similar behavior was observed in other measurements as well [53], which implies significant changes in electronic, magnetic and/or structural properties under pressure. In fact, the pressure dependence of $T_c$ exhibits an anomaly around 1 GPa as shown in Fig. 25. Investigation of the structural changes under high pressure gives direct information on the mechanism behind the enhancement of $T_c$ of FeSe under high pressure. Figure 26 summarizes the changes of the structural parameters under high pressure. In Fig. 26(c), which displays the pressure dependence of Se height from the Fe layer, a significant anomaly is observed around 1 GPa. To clarify the relationship between the Se height and $T_c$, we plotted the data together in Fig. 27 [57]. As indicated by an orange line in Fig. 27, the pressure dependence of $T_c$ has an anomaly at which the Se height suddenly drops, and FeSe shows a dramatic enhancement of $T_c$ up to 37 K. These results suggest that the $T_c$ seems to correlate strongly with the Se height. If this is so, what changes with the structural changes under high pressure? Nuclear magnetic resonance (NMR) studies indicate that antiferromagnetic fluctuations are enhanced together with the enhancement of $T_c$ under high pressure [56,58]. Figure 28 shows the temperature dependence of $1/T_1T$ for superconducting $Fe_{1.01}Se$ under high pressure up to 2.2 GPa, and for non-superconducting $Fe_{1.03}Se$ [58]. For a superconducting sample, $1/T_1T$ increases with decreasing temperature above $T_c$, which indicates that the superconductivity sets in at $T_c$ after the antiferromagnetic fluctuations are enhanced. Furthermore, $1/T_1T$ is enhanced by applying pressure, implying the huge pressure effect on $T_c$ in FeSe is directly associated with the enhancement of the antiferromagnetic fluctuations. From the above discussion, superconductivity in FeSe is likely to be correlated with the magnetic fluctuations. Furthermore, the change in local crystal structure is another important factor for explaining the enhancement of $T_c$ by applying pressure. In chapter 4, we extend the discussion on relationship between $T_c$ and anion height (P, As, Se, Te heights) to all Fe-based superconductors.

3-3. Pressure effect of FeTe

FeTe, one of the parent phases, shows antiferromagnetic ordering below 70 K. Generally antiferromagnetic ordering in the parent phase of Fe-based superconductor is suppressed by applying pressure. This in turn induces superconductivity, for example, in $SrFe_2As_2$ [52]. In view of this, it is expected that superconductivity would be induced in FeTe by applying pressure as in other Fe-based superconductors. Figure 29(a) shows the temperature dependence of resistivity for $Fe_{1.08}Te$ under high pressure [59]. A piston cylinder pressure cell, which can apply nearly hydrostatic pressure to the sample, was used in this measurement. With applying pressure, the sharp transition at 70 K, which corresponds to the structural and antiferromagnetic transition, was suppressed as indicated by red arrows. Blue arrows indicate the appearance of a high-pressure phase. A superconducting transition was not induced, presumably due to the appearance of the high-pressure phase above 1.5 GPa, as summarized in Fig. 29(b) [59- 61]. In contrast to FeAs-based compounds, superconductivity was not induced by the application of hydrostatic pressure for FeTe.

3-4. Pressure effects of FeTe-based superconductors

Here we summarize the pressure effects of $FeTe_{1-x}Se_x$ and $FeTe_{1-x}S_x$ superconductor. $FeTe_{1-x}Se_x$ shows a positive pressure effect on $T_c$ as in the case of FeSe [62-64]. Figure 30(a) shows the temperature dependence of resistivity for $Fe_{1.03}Te_{0.43}Se_{0.57}$ under high pressure. The pressure dependences of both $T_c$ and the crystal structure are summarized in Fig. 30(b). The $T_c$ reaches 23 K at ~2 GPa, and decreases under pressure above 2 GPa. While the low-temperature crystal structure below 2 GPa was orthorhombic as of FeSe, above 2 GPa it was monoclinic, which is the same as FeTe. In fact, the $T_c$ in FeTe$_{1-x}$Se$_x$ was enhanced under high pressure when the crystal structure was orthorhombic or tetragonal. In contrast, the $T_c$ decreases under high pressure when the crystal structure was monoclinic. These features are likely to be related to magnetism since the monoclinic phase in FeTe-based compounds has long-range antiferromagnetic ordering at ambient pressure.

Figure 31(a) shows the temperature dependence of resistivity for FeTe$_{1-x}$S$_x$ under high pressure [57]. The FeTe$_{0.8}$S$_{0.2}$ polycrystalline sample was exposed to air for several days to induce superconductivity as described in Ref. 40. The $T_c$ decreases with increasing pressure as plotted in Fig. 31(b). The negative pressure effect on $T_c$ is probably due to the appearance of a magnetism-related phase under high pressure as observed in FeTe$_{1-x}$Se$_x$. It is natural to think that superconductivity in FeTe$_{0.8}$S$_{0.2}$ is strongly suppressed because the composition is close to antiferromagnetic FeTe, in other words, it is located close to the boundary between an antiferromagnet and superconductor as discussed in section 2-5. However, Zhang et al. has reported a possibility of an enhancement of $T_c$ for the FeTe$_{1-x}$S$_x$ sample containing inhomogeneity of composition [65]. To clarify the intrinsic properties of FeTe$_{1-x}$S$_x$, a detailed pressure study using a high-quality FeTe$_{1-x}$S$_x$ sample that shows bulk superconductivity is necessary.

The pressure effect on $T_c$ for Fe-chalcogenide superconductors is summarized in Fig. 32. FeSe shows the largest pressure effect, and the $T_c$ of 37 K at 4-6 GPa is the highest among the Fe chalcogenides. Although FeSe has an anomaly in the pressure-$T_c$ curve, the low-temperature crystal structure does not change up to ~6 GPa. In contrast, the FeTe-based superconductors exhibit the pressure-induced structural change to the monoclinic structure, which is the same as in the antiferromagnetically ordered FeTe. In fact, the pressure effect tends to be suppressed when the composition approaches that to FeTe. The most important factor for achieving a high $T_c$ is to maintain the orthorhombic or tetragonal structure under high pressure.

3-5. Effect of strain on $T_c$

Since the superconducting properties of Fe chalcogenide are strongly affected by the crystal structure, a higher $T_c$ could be achieved at an ambient pressure by applying internal strain. A thin film is a candidate for exhibiting a higher $T_c$ at an ambient pressure, because strain stress can be generated by changing the lattice constant of the substrate and/or the thickness of the film. At present, the superconducting thin films of FeSe, FeTe$_{1-x}$Se$_x$ and FeTe$_{1-x}$S$_x$ have already been successfully fabricated [66-72]. In fact, an enhancement of $T_c$ by changing the thickness of the FeTe$_{1-x}$Se$_x$ film, which in turn changes the lattice constants, has been reported [70]. Figure 33(a) shows the temperature dependence of resistivity for a FeTe$_{0.5}$Se$_{0.5}$ film. The $T_c$ increases up to 21 K, which is significantly higher than that of the bulk sample. The thickness dependence of the lattice constant and $T_c$ is summarized in Fig. 33(b) and (c). When the thickness is 200 nm, the lattice constant becomes the minimal value, and the $T_c$ reaches the highest. The $T_c$ of 21 K in FeTe$_{1-x}$Se$_x$ is almost the same value as the highest $T_c$ under high pressure, $T_c$ = 23 ~ 25 K [62,63]. From above, we expect to produce a FeSe thin film with a $T_c$ above 30 K, because the $T_c$ of FeSe reaches 37 K under high pressure.

The strain-stress effect on $T_c$ has been reported in the FeSe polycrystalline sample as well [73,74]. Figure 34 shows the temperature dependence of magnetic susceptibilities for as-grown and annealed FeSe samples [73]. The inset shows the temperature dependence of resistivity for the same samples. Although the $T_c$ of the as-grown sample is higher than 10 K, annealing decreases the $T_c$, makes the transition sharper and stabilizes the PbO-type structure. Furthermore, in Ref. 74, a $T_c$ near 23 K has been reported in a strain-stressed FeSe single crystal. If a FeSe bulk sample having a $T_c$ near 37 K at ambient pressure can be obtained, it will be a very good candidate for applications such as superconducting wires.

## 4. Mechanism of superconductivity in Fe chalcogenide

### 4-1. Anion height dependence of $T_c$

As discussed above, the enhancement of $T_c$ in FeSe is strongly correlated with the Se height from the Fe-square plane. For this reason, we discuss the anion height dependence of $T_c$ for typical Fe-based superconductors including FeP-, FeAs-based and Fe-chalcogenide superconductors. Figure 35 shows the anion height dependence of $T_c$ for the typical Fe-based superconductors [57,75]. The data was selected according to using two following criteria below: (1)the valence of Fe of the superconductor should be close to +2. (2)the $T_c$ of the superconductor should be the highest within the system. The second condition demands that the degree of doping or applied pressure is optimal. This is because that superconductivity occurs when the structural and antiferromagnetic transition are suppressed by elemental substitution and/or applying external pressure for almost all the Fe-based superconductors.

The anion height increases with changing anions as an order of P, As, Se and Te. Surprisingly, all data, both those under applied and at ambient pressure, obey a symmetric curve with a peak around 1.38 Å. This suggests that the $T_c$ in Fe-based superconductors can be simply explained by the structural parameter of anion height, irrespective of the difference in the anion that composes the superconducting layers. As discussed above, the pressure dependence of $T_c$ for FeSe exhibits an anomaly around 1-2 GPa; the $T_c$ is enhanced up to 37 K above 2 GPa. As plotted in Fig. 35, the data point of FeSe at ambient pressure does not obey the curve. However, the data points approach the unique curve with increasing pressure, and fall onto it above 2 GPa. In this sense, one can say that the "intrinsic" or optimal superconductivity of FeSe is induced by applying pressure above 2 GPa. In contrast, the data points for either $FeTe_{1-x}Se_x$ or $FeTe_{1-x}S_x$ under high pressure do not follow the unique curve. The difference between the FeTe-based superconductors ($FeTe_{1-x}Se_x$ and $FeTe_{1-x}S_x$) and the other Fe-based superconductors, i.e., FeSe, FeP- and FeAs-based superconductors, lies in whether disorder at the anion site exists or not. As discussed here, the $T_c$ is sensitive to changes in the anion height. Therefore, the changes in the superconducting properties of FeTe-based compounds under pressure would not be easily understood. In fact, the existence of the disorder at the anion site has been detected using synchrotron x-ray diffraction and extended x-ray absorption fine structure (EXAFS) [76-78]. The anion Se and Te in $FeTe_{1-x}Se_x$ each take on different but universal (sample independent) height; the difference in the anion height between Se and Te was determined to be 0.24 Å for $FeTe_{0.5}Se_{0.5}$. We have used the average value of the heights in Fig. 35.

The relationship between the anion height and the superconducting properties has been taken up from a theoretical point of view [79-81]. Kuroki et al. argue that the pairing symmetry of Fe-based superconductivity changes from nodal low-$T_c$ paring to nodeless high-$T_c$ paring as the anion height is increased [79]. Moon and Choi discussed the correlation between anion height and magnetism in Fe chalcogenide, and suggested that the relatively high anion height is responsible for the unique magnetism in FeTe [80]. Anion height is thus key parameter that affects the intrinsic properties of Fe-based superconductors.

4-2. Superconducting gap observation in Fe chalcogenide

Experimental information on the superconducting gap is of central in discussing the physical properties of the superconducting state. Photoemission spectroscopy and scanning tunnel spectroscopy (STS), have been employed to investigate the electronic state and superconducting gap of Fe chalcogenides [82-86]. This led, in particular, to the successful observation of the superconducting gap of FeTe$_{1-x}$Se$_x$ [84-86], owing to the availability of the large high-quality single crystals. Figure 36 shows the superconducting gap for Fe$_{1.05}$Te$_{0.85}$Se$_{0.15}$ obtained using STS [85]. The value of $2\Delta/k_BT_c$ estimated from the superconducting gap shows a larger value of 6-7, which exceeds the value predicted by the BCS theory. This suggests the strong-coupling nature of superconductivity in FeTe$_{1-x}$Se$_x$. The strong-coupling nature was also suggested from the high-field transport measurement [87]. Figure 37 shows the field-temperature phase diagram for FeSe and FeTe$_{1-x}$Se$_x$. The temperature dependence of the upper critical field for FeTe$_{1-x}$Se$_x$ deviates from the WHH theory [88] due to the Pauli paramagnetic effect and a large $2\Delta/k_BT_c$ value is obtained, while FeSe agree with the WHH theory. The strong-coupling nature of superconductivity was also reported for BaFe$_2$As$_2$. These results may be relevant to the mechanism of Fe-based superconductivity.

Recently, Hanaguri et al. has reported a scanning tunneling microscopy (STM) study on FeTe$_{1-x}$Se$_x$ and showed an evidence for s$_\pm$-wave superconductivity, which had been predicted by theoretical studies [86,89,90]. The authors performed spectroscopic-imaging STM (SI-STM) on single crystals of FeTe$_{1-x}$Se$_x$. Figure 38(a) and (b) are schematic images of the reciprocal-space electronic states of the Fe-based superconductor; (a) shows the Fermi surface and inter -Fermi-pocket scatterings in **k** space, and (b) shows the expected quasi-particle interference (QPI) spots in **q** space associated with the inter-Fermi-pocket scatterings. Theoretically, sign-preserving scatterings indicated by blue filled circles should be enhanced by a magnetic field, whereas sign-reversing scatterings should be suppressed for the s$_\pm$-wave superconducting gap. Figure 38(c) shows the experimental result of magnetic field-induced changes in the QPI intensities. A clear difference in signals at **q**$_2$ and **q**$_3$ was observed, which indicates that the superconducting gap of FeTe$_{1-x}$Se$_x$ possesses s$_\pm$-wave symmetry.

5. Research for application of Fe chalcogenide

5-1. Superconducting wire fabrication

Superconducting wire is used in a high-field magnet since it can sustain a high electrical current flow using the zero-resistivity state. The upper critical field value is a critical factor in selecting the material for a superconducting wire, which is the building block of high-field magnet. Fe-based superconductors, which generally have high upper critical fields, are therefore potentially suitable for this purpose. Fe-chalcogenide superconducting wires have been fabricated using both in-situ and ex-situ methods [91,92]. Figure 39(a) shows the temperature dependence of resistivity for an $FeTe_{0.5}Se_{0.5}$ superconducting wire. These wires were fabricated with an Fe sheath using an ex-situ powder-in-tube method and post-annealed in evacuated quartz tubes. An image of the cross section is shown in the inset of Fig. 39(b). The connectivity between the superconductor and the sheath seems to be good. Zero resistivity was observed for the as-fabricated tube. When annealed around 200 ºC, the superconducting transition becomes sharper and $T_c$ increases. Figure 39 (b) shows the magnetic field dependence of critical current density ($J_c$) at 4.2 K. Although the obtained $J_c$ (0 T) is ~ 40 $A/cm^2$, which is still smaller than the value used in existing applications, it can be improved by optimization of the annealing process, an introduction of pinning centers, and fabrication of the multi-core wire. If the $J_c$ value becomes high enough for application, it should be useful for a high-field magnet because Fe-chalcogenide superconductors have a high $H_{c2}$.

5-2. Thin film

Thin film fabrication is an important technique for application of superconductivity. Several Fe-chalcogenide thin films have already been fabricated [66-72]. In ref. 66, the thickness dependence of the transition temperature in FeSe films prepared at a low temperature was reported. Superconductivity appears when the thickness exceeds 140 nm, and the $T_c$ increases with increasing thickness. This reflects the sensitivity of superconductivity to the crystal structure as discussed in chapters 3 and 4. Lattice distortions caused by the mismatch in the lattice constants between the substrate and Fe chalcogenides would affect the superconducting properties when the thickness of the film is small.

Nanosheets of Fe chalcogenides were fabricated using a chemical process [93]. The TEM images of FeSe nanosheet are shown in Fig. 40(a-c). The x-ray diffraction pattern (d) and the selected area electron diffraction (e) indicate that the nanosheets possess the tetragonal structure and are aligned along the $c$ axis. Nanosheets of FeTe and $FeTe_{1-x}Se_x$ were synthesized as well. Although the Fe-chalcogenide nanosheets with the PbO-type structure have been successfully obtained, superconductivity was not observed. To clarify the factors important for superconductivity in Fe chalcogenides, more detailed studies on thin film and nanosheet are needed. Recently, we found that the scotch-tape method, which has been used in the research on graphene and other layered materials, is applicable to Fe-chalcogenide single crystals. The single crystals cleaved using the scotch-tape method were placed on a substrate. Figure 41(a) is an optical-microscope image of an FeTe$_{0.936}$S$_{0.064}$ single crystal cleaved and placed on the Si substrate using the scotch-tape method. Figure 41(b) is an atomic force microscopy (AFM) image of the single crystals on the Si substrate. The thickness of the cleaved crystals was estimated to be 140 nm for A, 57 nm for B, 40 nm for C and 48 nm for D. The thickness of 40 nm roughly corresponds to the thickness of 60 - 70 layers. Investigation of the thickness dependence of physical properties would be an interesting direction to pursue in the field of Fe-based superconductivity.

6. Summary

PbO-type FeSe shows superconductivity with a $T_c \sim 10$ K, and has been drawing much attention as the simplest Fe-based superconductor. The $T_c$ of FeSe dramatically increases with applied pressure, and reaches 37 K at 4 – 6 GPa. The enhancement of $T_c$ is correlated with the enhancement of antiferromagnetic fluctuations and/or a change in the anion height. The anion height dependence of $T_c$ is expected to be applicable to all Fe-based superconductors.

PbO-type FeTe, the parent compound of Fe-based superconductor, exhibits antiferromagnetic ordering below 70 K. Partial substitution of Te by Se suppresses antiferromagnetism and induces superconductivity. Since large and high-quality FeTe$_{1-x}$Se$_x$ single crystals can be grown, many studies have been performed using the crystals. Among them, perhaps the most significant result is the STM study on FeTe$_{1-x}$Se$_x$ single crystals by Hanaguri et al., which shows clear evidence of s$_\pm$-wave symmetry, which was theoretically predicted as the common symmetry of Fe-based superconductors. Partial substitution of Te by S also suppresses the antiferromagnetism and induces superconductivity. However, the solubility limit of S for the Te site is low due to a large difference in the ionic radius between S$^{2-}$ and Te$^{2-}$. Therefore, as-grown samples do not exhibit bulk superconductivity. Rather surprisingly, the superconducting properties are improved by post treatments, for example, air exposure, oxygen annealing, immersion into hot water and alcoholic liquids, by which bulk superconductivity is induced. Oxygen thus seems to be the key ingredient behind the appearance of bulk superconductivity in FeTe$_{1-x}$S$_x$.

Fe chalcogenide superconductors are good candidates for applications since the structure and composition is the simplest among the Fe-based superconductors. Since Fe-chalcogenide superconductors have high upper critical fields, using these materials to produce superconducting wires is an important application and fabrication challenge. Fabrication of thin films and nanosheets will advance the understanding on the intrinsic properties of Fe chalcogenide, and at the same time will open up a new field for applications using Fe-chalcogenide superconductors.

Acknowledgements

We thank the collaborators, Prof. H. Kotegawa (Kobe University), Prof. H. Tou (Kobe University), Prof. T. Yokoya (Okayama University), Dr. T. Kida (Osaka University), Prof. M. Hagiwara (Osaka University), Dr. T. Kato (Tokyo University of Science), Prof. H. Sakata (Tokyo University of Science), Dr. H. Okada (Tohoku Gakuin University), Prof. H. Takahashi (Nihon University), Prof. N. L. Saini (University of Roma), Prof. S. Margadonna (University of Edinburgh), Prof. K. Prassides (University of Durham) and Prof. H. Kumakura (National Institute for Materials Science). The authors thank the group members (National Institute for Materials Science), Mr. Y. Kawasaki, Mr. K. Deguchi, Mr. T. Watanabe, Dr. T. Ozaki, Dr. S. Tsuda and Dr. T. Yamaguchi for useful discussions and experimental help. This work was partly supported by Grant-in-Aid for Scientific Research (KAKENHI).

Figure captions

Fig. 1. Crystal structure of typical Fe-based superconductors. (a) LaFeAsO. (b) BaFe$_2$As$_2$. (c) FeSe. (d) *ab* plane of FeSe. The Fe-square lattice is a common structure of Fe-based superconductor. The images were drawn using VESTA [94].

Fig. 2. Phase diagram of Fe$_{1+d}$Se [20].

Fig. 3. Temperature dependence of lattice constants for FeSe [21].

Fig. 4. Temperature dependence of resistivity for FeSe$_{0.88}$. The inset shows the normalized temperature dependence of upper critical field [11].

Fig. 5. Temperature dependence of magnetic susceptibility for Fe$_{1+d}$Se [20].

Fig. 6. (a) Crystal structure of Fe(Te,Se). Magnetic structures of (b) FeTe and (c) BaFe$_2$As$_2$ are shown in the primitive Fe square lattice for comparison. Note that the basal square lattice of the PbO unit cell in (a) is $\sqrt{2} \times \sqrt{2}$ superlattice of that in (b). (d),(e) The magnetic Bragg peak ($\delta$, 0, 1/2) (blue symbols) and the splitting of the structural peak (200) or (112) of the tetragonal phase (red symbols) show the thermal hysteresis in the first-order transition [26].

Fig. 7. Temperature dependence of resistivity for FeSe and FeTe.

Fig. 8. $^{57}$Fe Mössbauer spectra of Fe$_{1.08}$Te at room temperature, 77 and 4.2 K. The spectrum at room temperature is fitted by two types of doublets, which are attributed to the two Fe sites. One site is the Fe in the FeTe layer (Fe-1 site), and the other is Fe which exists at the interlayer site (Fe-2 site). At 4.2 K, the clear magnetic sextet corresponding to the magnetic ordering around 70 K was observed [23].

Fig. 9. Schematic in-plane spin structure of Fe$_{1.068}$Te and FeAs-based SrFe$_2$As$_2$. The solid arrows and hollow arrows represent two sublattices of spins, which can be either parallel or anti-parallel. The shaded area indicates the magnetic unit cell [25].

Fig. 10. Temperature dependence of magnetic susceptibility for FeTe$_{1-x}$Se$_x$ single crystals.

Fig. 11. (a) Temperature dependence of both resistivity and magnetic susceptibility for FeTe$_{0.5}$Se$_{0.5}$ single crystal. (b) Photograph of Fe$_{1.13}$Te$_{0.73}$Se$_{0.27}$ single crystals [32].

Fig. 12. Phase diagram of FeTe$_{1-x}$Se$_x$.

Fig. 13. (a-f) Temperature dependence of magnetic states for FeTe$_{1-x}$Se$_x$. (g) Phase diagram of Fe$_{1.03}$Te$_{1-x}$Se$_x$ determined from (a)-(f) [35].

Fig. 14. (a) Temperature dependence of magnetic susceptibility for Fe$_{1.1}$Te$_{1-x}$Se$_x$. The inset shows both ZFC and FC susceptibility. (b) Se concentration dependence of magnetic transition temperature and hyper fine field ($H_{hf}$) [37].

Fig. 15. The energy scan at Q = (0.46, 0.46, 0.66) as a function of temperature indicating association between the spin resonance and superconductivity in FeSe$_{0.4}$Te$_{0.6}$. The sample turned background was subtracted from the data. The inset shows the integrated intensity of the resonance between 5 meV and 8 meV as a function of temperature, and the line is a fit to mean field theory with $T_c$ = 14 K [38].

Fig. 16. Temperature dependence of resistivity for FeTe$_{1-x}$S$_x$ [19].

Fig. 17. Starting nominal S concentration dependence of lattice constant $c$ for FeTe$_{1-x}$S$_x$ synthesized by 3 methods.

Fig. 18. (a) Temperature dependence of resistivity with several air-exposure time for FeTe$_{0.8}$S$_{0.2}$ polycrystalline sample prepared by the solid-state reaction method. (b) Temperature dependence of magnetic susceptibility with several air-exposure time for FeTe$_{0.8}$S$_{0.2}$ polycrystalline sample [40].

Fig. 19. Superconducting volume fraction estimated from the susceptibility measurement for liquid-immersed FeTe$_{0.8}$S$_{0.2}$ polycrystalline sample. Superconductivity is induced by immersing the sample into water-ethanol mixture and alcoholic drinks. The value of superconducting volume fraction is plotted as a function of alcoholic concentration of the liquid [42].

Fig. 20. (a) Temperature dependence of resistivity for as-grown FeTe$_{0.8}$S$_{0.2}$ and that after oxygen annealing for 2 hours at 200 ºC. (b) Temperature dependence of magnetic susceptibility for FeTe$_{0.8}$S$_{0.2}$ polycrystalline sample annealed for 2 hours at 100, 200, 300 and 400 ºC [43].

Fig. 21. (a) Observed (1 0 1) reflection in the x-ray diffraction pattern for FeTe$_{0.8}$S$_{0.2}$ polycrystalline sample annealed for 2 hours at 100, 200, 300 and 400 ºC. (b,c) lattice constants $a$ and $c$ for oxygen-annealed FeTe$_{0.8}$S$_{0.2}$. For comparison, the lattice constants of FeTe$_{0.8}$S$_{0.2}$ prepared using the melting method are plotted [43].

Fig. 22. (a) Temperature dependence of resistivity for as-grown FeTe$_{1-x}$S$_x$ single crystals. (b) Temperature dependence of resistivity for FeTe$_{1-x}$S$_x$ single crystals annealed in oxygen for 2 hours at 200 °C [44].

Fig. 23. Phase diagram of FeTe$_{1-x}$S$_x$ single crystals grown using the self-flux method. Due to the low solubility limit of S for the Te site, single crystals with $x > 0.13$ cannot be obtained [44].

Fig. 24. Temperature dependence of resistivity for FeSe under high pressure [56].

Fig. 25. Pressure dependence of $T_c$ for FeSe.

Fig. 26. (a-e)Pressure dependence of crystal structural parameters for FeSe. (a) Fe-Se distance. (b) Se-Fe-Se angle. (c) Se height and interlayer Se-Se distance. (d) Lattice constants. (e) Volume of lattice. The inset shows volume fraction of tetragonal phase. (f) Schematic images of the structural parameters [54].

Fig. 27. Pressure dependence of $T_c$ and Se height for FeSe. An anomaly is observed at 1 – 2 GPa in both $T_c$ and Se height [57].

Fig. 28. Temperature dependence of $1/T_1T$ for Fe$_{1.01}$Se under high pressure. The inset shows the temperature dependence of $1/T_1$ for Fe$_{1.01}$Se under high pressure [58].

Fig. 29. (a) Temperature dependence of resistivity for Fe$_{1.08}$Te under high pressure. (b) Pressure-temperature phase diagram of Fe$_{1.08}$Te [59].

Fig. 30. (a) Temperature dependence of resistivity for Fe$_{1.03}$Te$_{0.43}$Se$_{0.57}$ under high pressure. (b) Pressure-temperature phase diagram of Fe$_{1.03}$Te$_{0.43}$Se$_{0.57}$ [62].

Fig. 31. Temperature dependence of resistivity for air-exposed FeTe$_{0.8}$S$_{0.2}$ polycrystalline sample under high pressure. (b) Pressure-temperature phase diagram of air-exposed FeTe$_{0.8}$S$_{0.2}$ [57].

Fig. 32. Pressure dependence of $T_c$ for FeSe, FeTe$_{1-x}$Se$_x$ and FeTe$_{1-x}$S$_x$.

Fig. 33. (a) Temperature dependence of resistivity for FeTe$_{0.5}$Se$_{0.5}$ thin film with several thicknesses.

(b) Thickness dependence of lattice constant $a$ and $T_c$ for FeTe$_{0.5}$Se$_{0.5}$ thin film [70].

Fig. 34. Temperature dependence of magnetic susceptibility for as-grown and annealed Fe$_{1.03}$Se polycrystalline samples. The inset shows the temperature dependence of resistivity for as-grown and annealed Fe$_{1.03}$Se polycrystalline samples [73].

Fig. 35. Anion height dependence of $T_c$ for the typical Fe-based superconductors. Filled and open symbols indicate the data points at ambient pressure and under high pressure, respectively [57].

Fig. 36. Spatially averaged spectrum at 4.2 K normalized to the background conductance for Fe$_{1.05}$Te$_{0.85}$Se$_{0.15}$. The curve indicates a fit of the calculated density of state for an $s$-wave superconductor to the data [85].

Fig. 37. Field-temperature phase diagram for FeSe and FeTe$_{0.75}$Se$_{0.25}$. Solid lines are the fitting curve using the WHH theory. The dashed line is a fitting curve with a large Maki parameter [87].

Fig. 38. (a,b) Schematic of reciprocal-space electronic states of an iron-based superconductor. (a) Fermi surface and inter-Fermi-pocket scatterings in k space. There are two hole cylinders and two electron cylinders around G and M points, respectively. For simplicity, two Fermi cylinders are represented by one circle. The black dotted diamond denotes the crystallographic Brillouin zone, and the red square indicates the "unfolded" Brillouin zone defined by the unit cell containing one iron atom. Signs of the SC gap expected for $s_\pm$-wave symmetry are shown by different colors. The arrows denote inter-Fermi-pocket scatterings. $q_2$ and $q_3$ connect different pockets, whereas $q_1$ is an umklapp process. (b) Expected QPI spots in q space associated with inter-Fermi-pocket scatterings. Blue filled circles and red filled diamonds represent sign-preserving and sign-reversing scatterings, respectively, for the $s_\pm$-wave SC gap. The former will be enhanced by a magnetic field, whereas the latter may be suppressed. (c) Magnetic field–induced change in QPI intensities indicates the $s_\pm$-wave symmetry.. It is clear that signals at $q_2$ and $q_3$ exhibit totally opposite behavior, indicating that these two scatterings have different characters. This pattern can only be explained if the SC gap possesses $s_\pm$-wave symmetry [86].

Fig. 39. (a) Temperature dependence of resistivity for FeTe$_{0.5}$Se$_{0.5}$ wire fabricated using ex-situ powder-in-tube method with various post annealing. (b) Magnetic field dependence of critical current density at 4.2 K for FeTe$_{0.5}$Se$_{0.5}$ wire. The inset is a optical microscope image of the cross section of the wire [92].

Fig. 40. (a-c) TEM images of FeSe nanosheets. (d) X-ray diffraction pattern for FeSe nanosheets. (e) Selected area electron diffraction pattern for FeSe nanosheet.

Fig. 41. (a) Optical microscope image of FeTe$_{0.936}$S$_{0.064}$ single crystals cleaved and left on Si substrate using the scotch tape method. (b) AFM image of the sheet of FeTe$_{0.936}$S$_{0.064}$ single crystals on Si substrate. The thickness of the sheets A, B, C and D is 140, 57, 40 and 48 nm, respectively.

Figures

Fig. 1

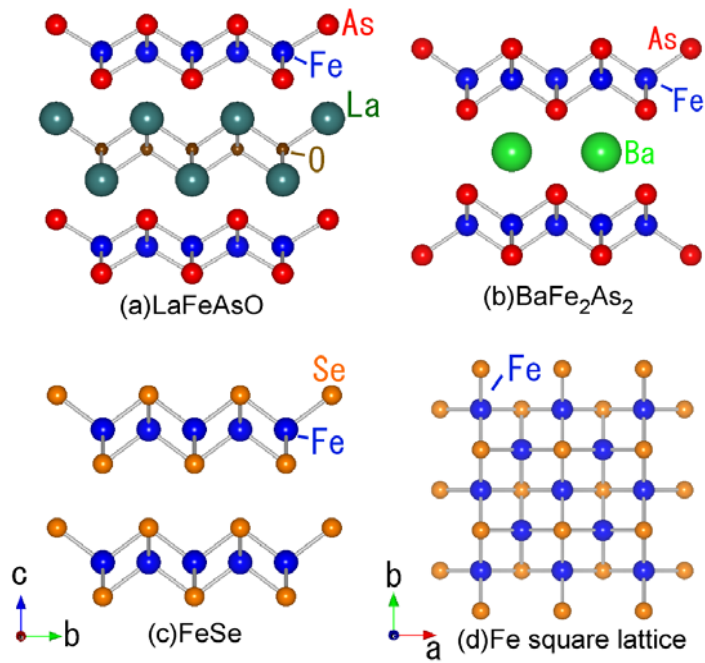

Fig. 2

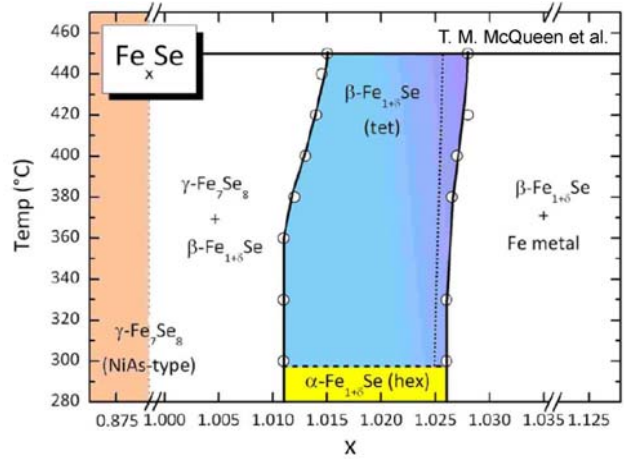

Fig. 3

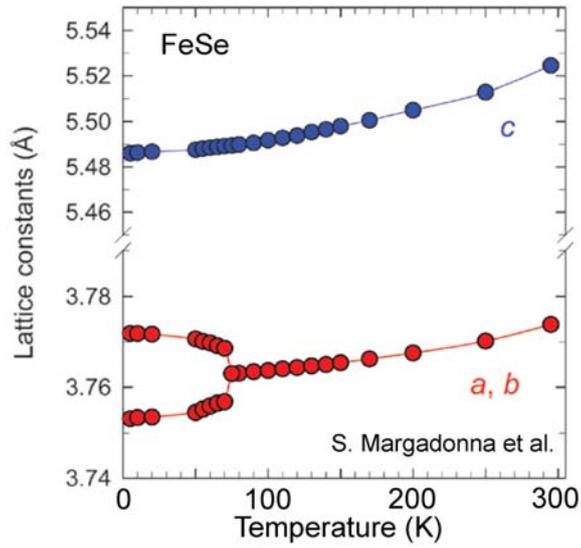

Fig. 4

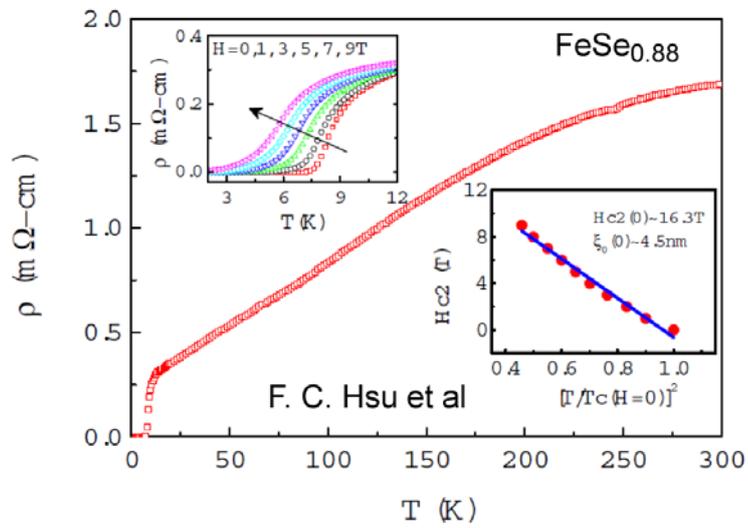

Fig.5

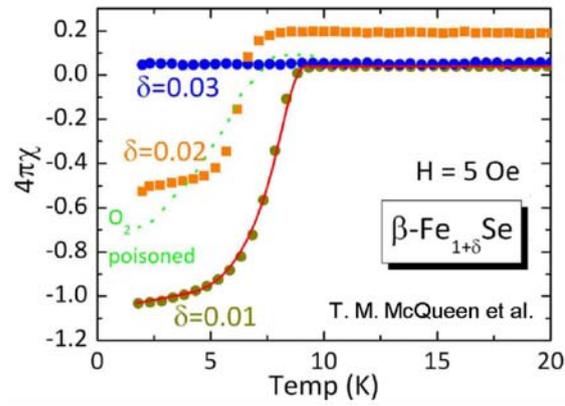

Fig. 6

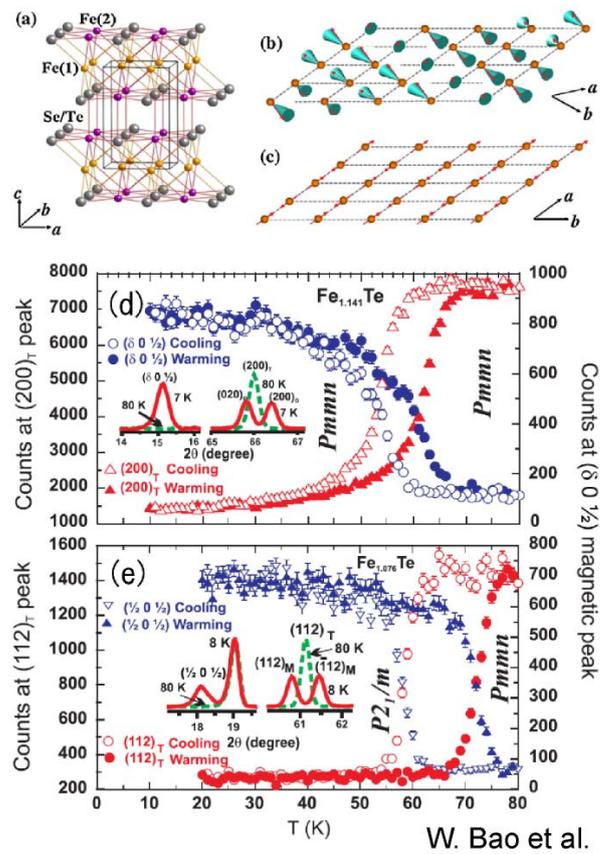

Fig. 7

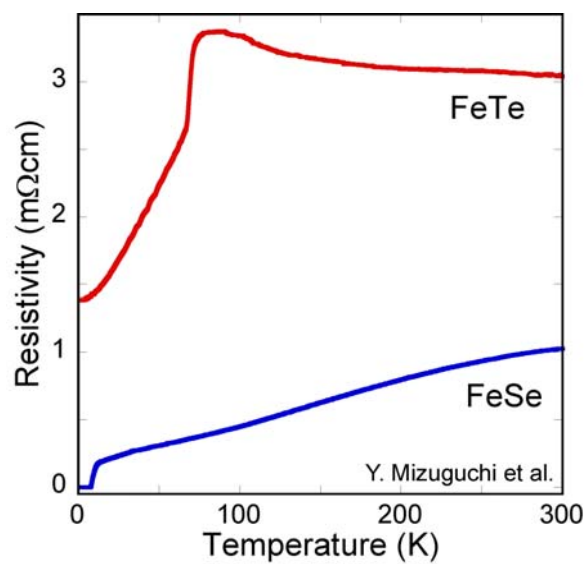

Fig. 8

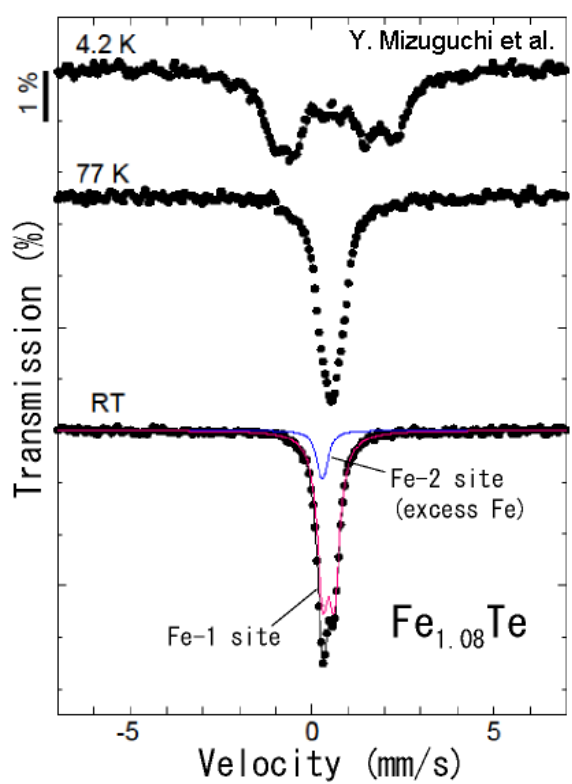

Fig. 9

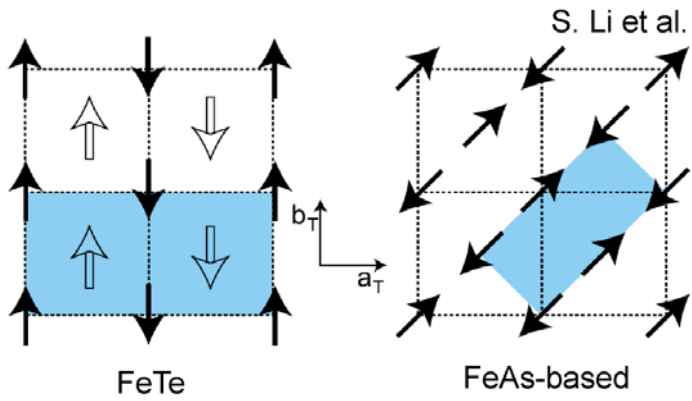

Fig. 10

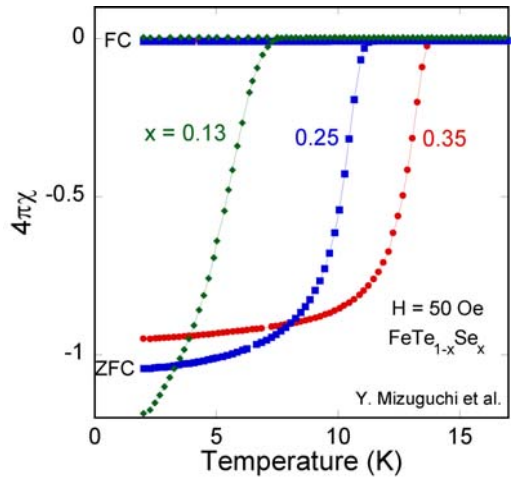

Fig. 11

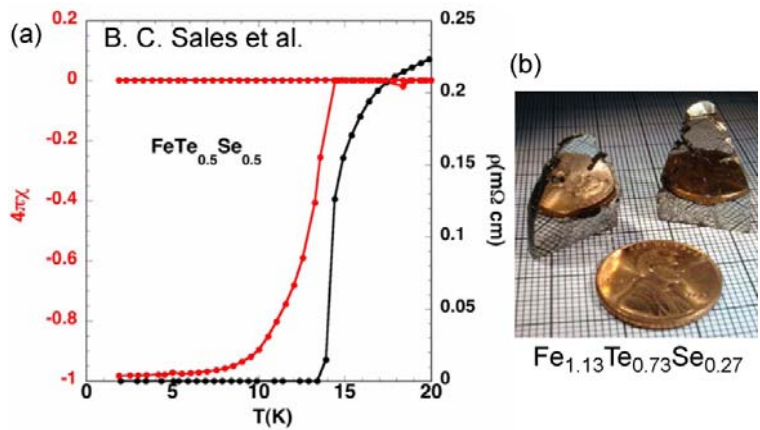

Fig. 12

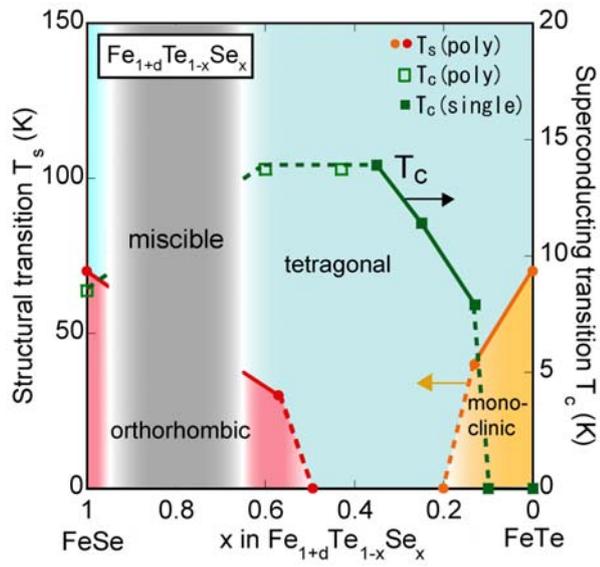

Fig. 13

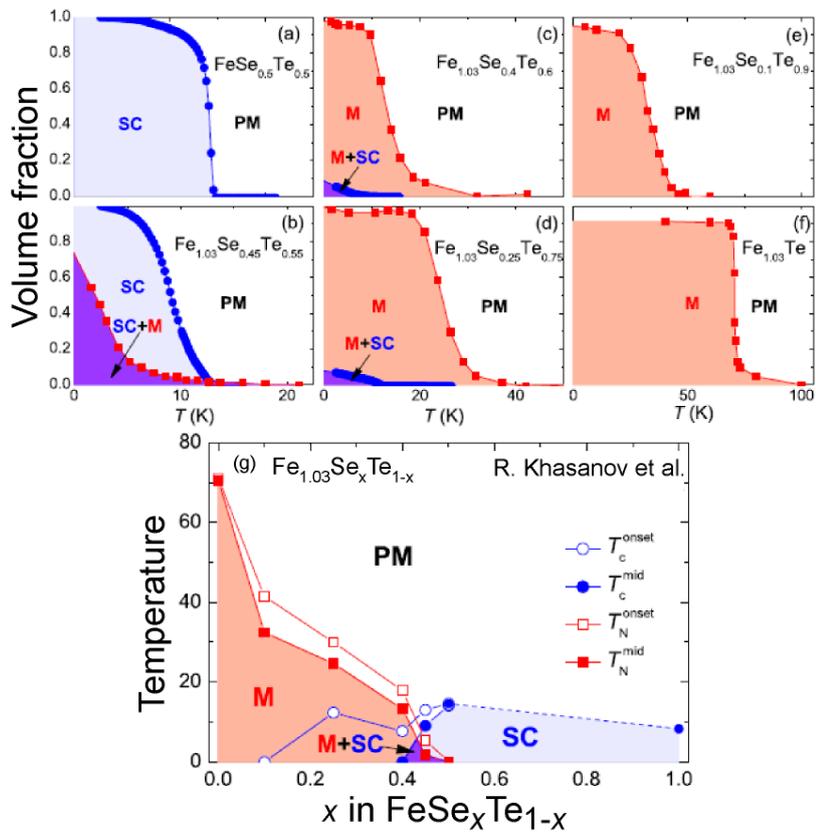

Fig. 14

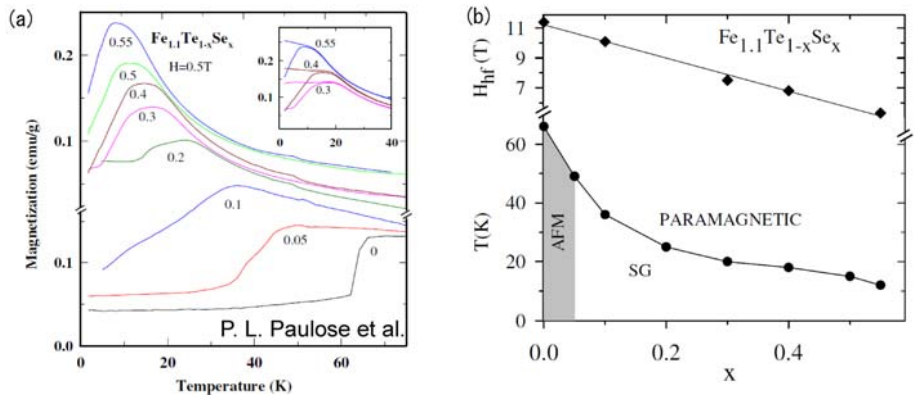

Fig. 15

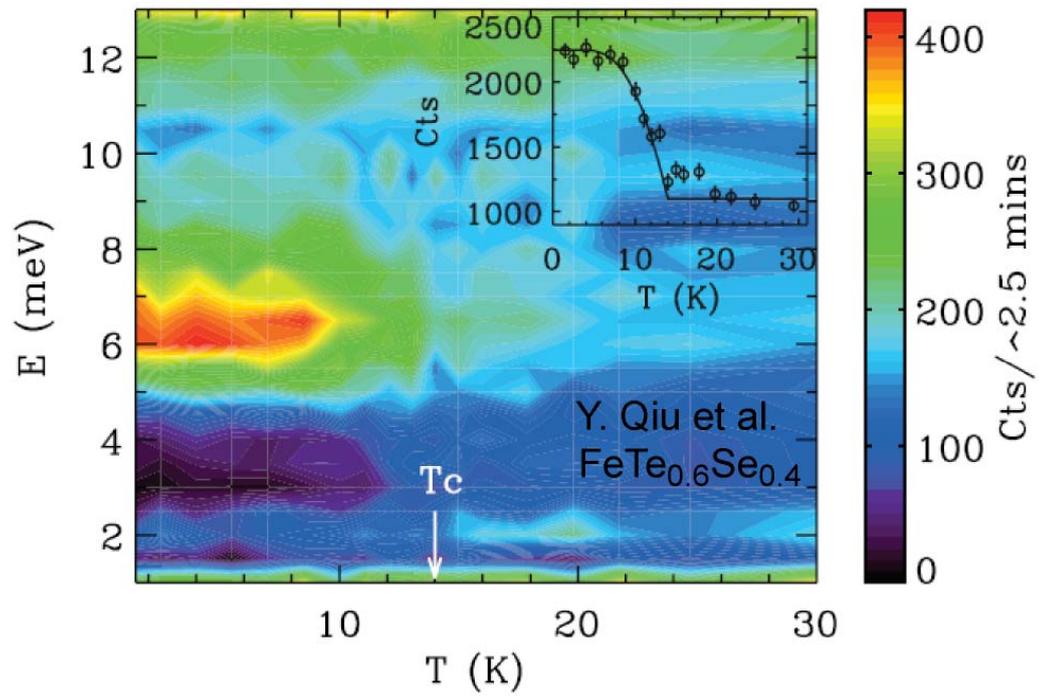

Fig. 16

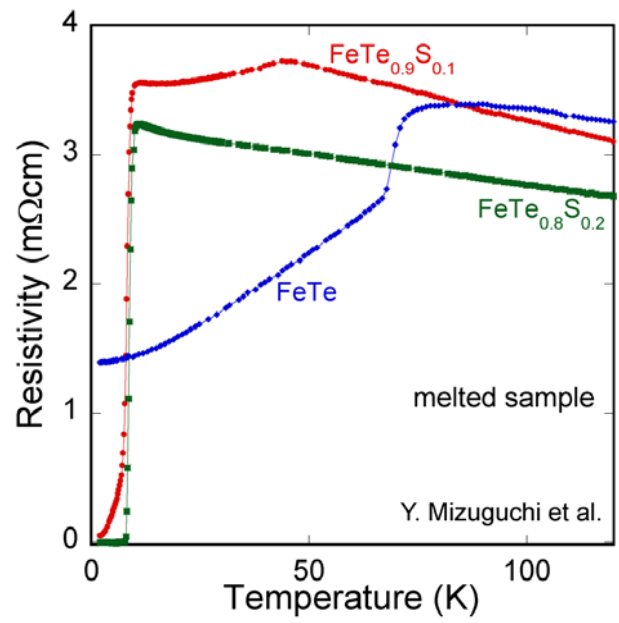

Fig. 17

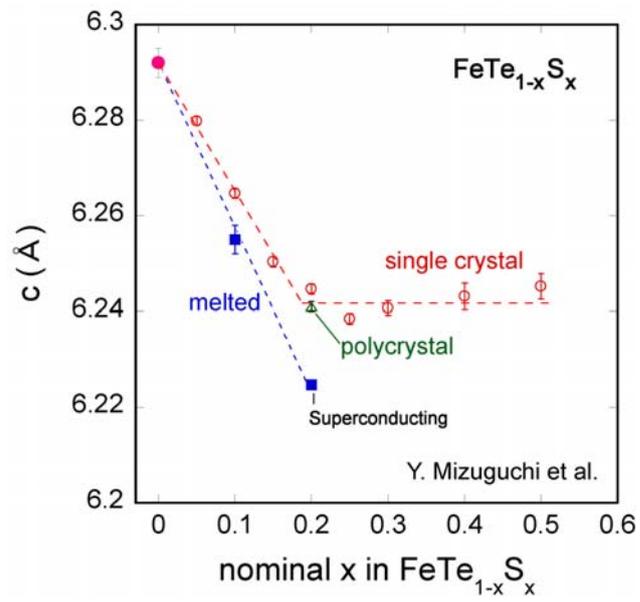

Fig. 18

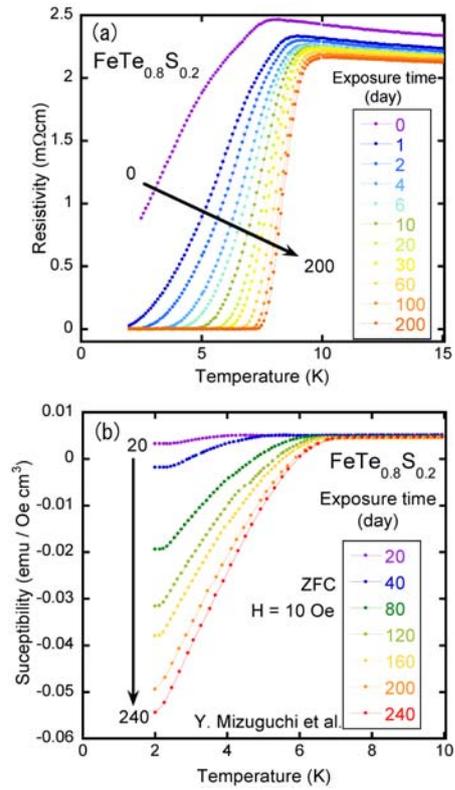

Fig. 19

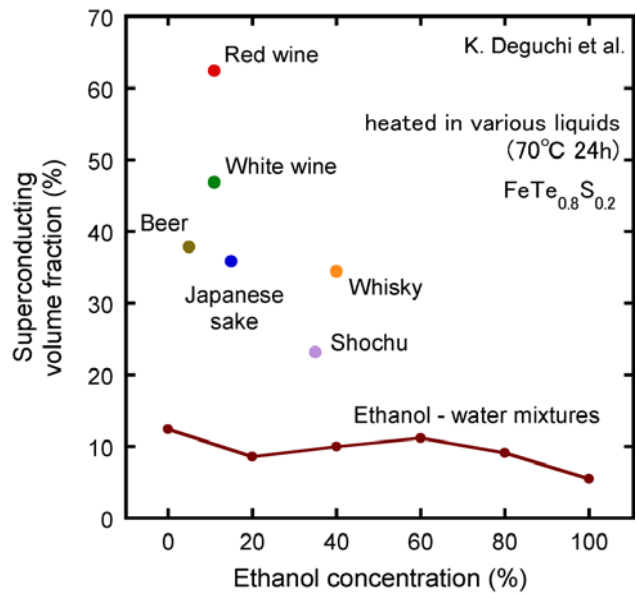

Fig. 20

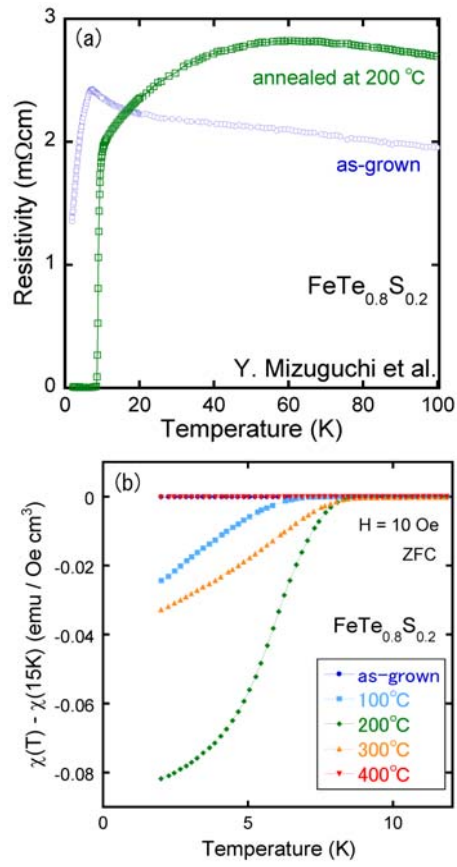

Fig. 21

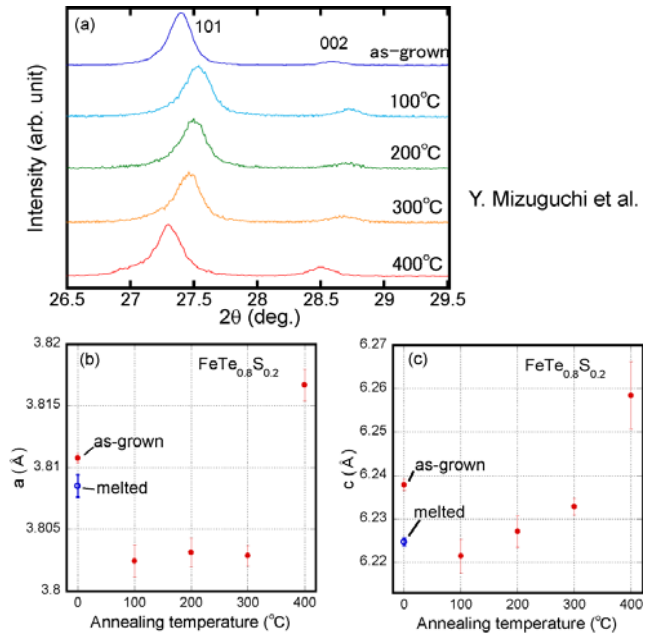

Fig. 22

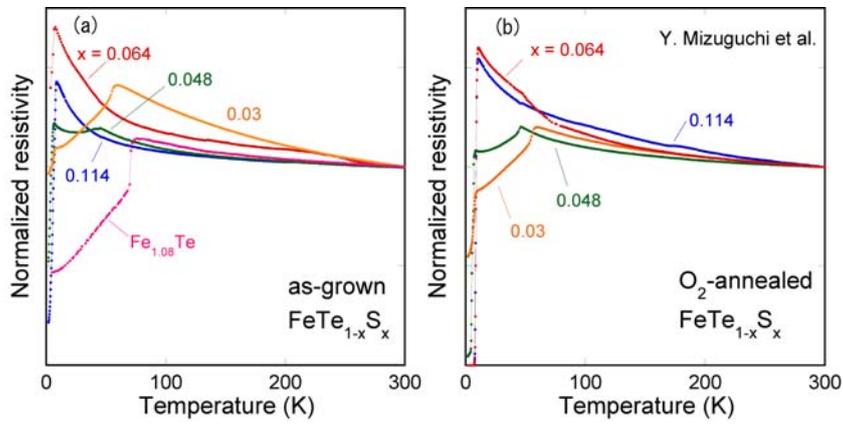

Fig. 23

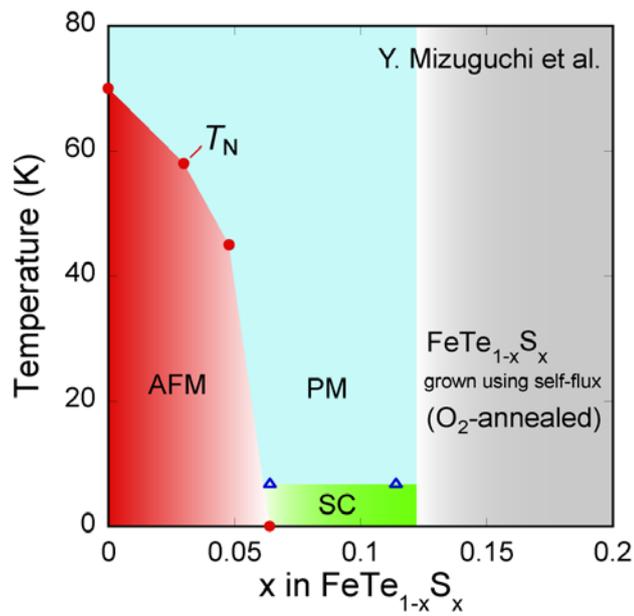

Fig. 24

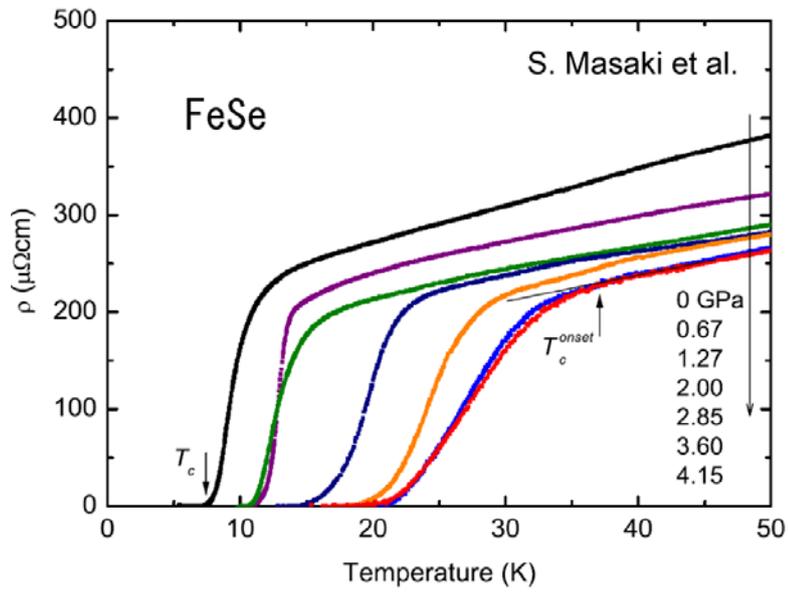

Fig. 25

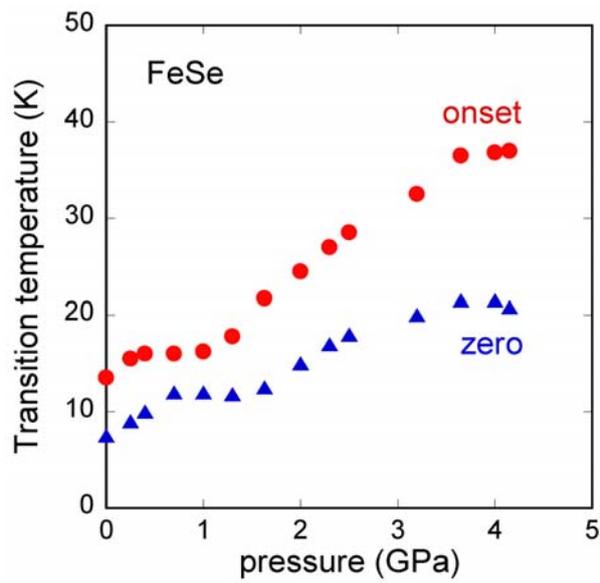

Fig. 26

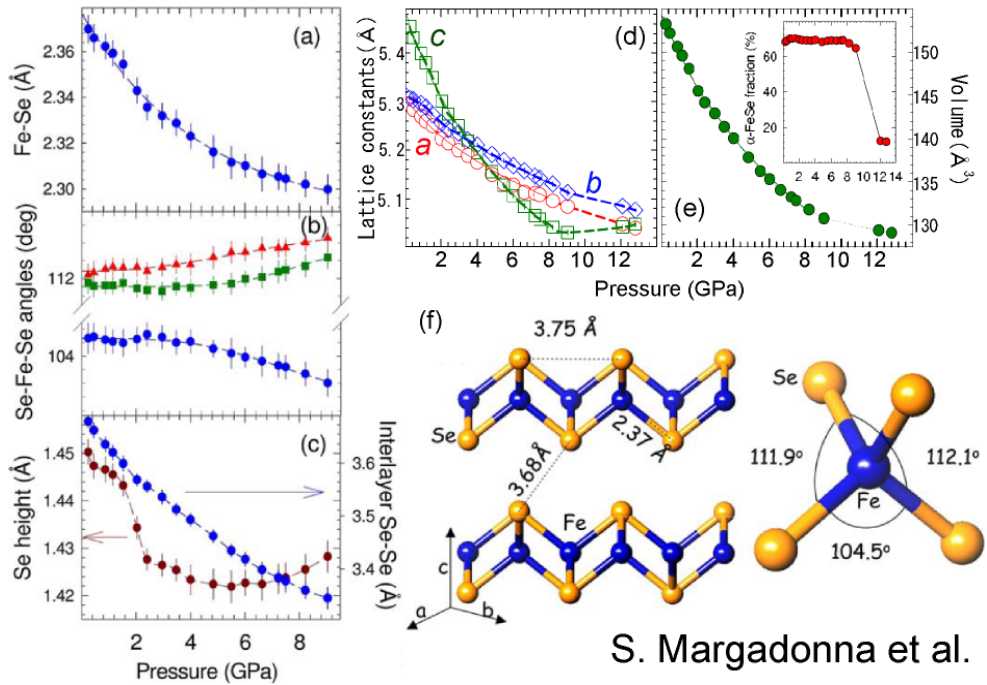

S. Margadonna et al.

Fig. 27

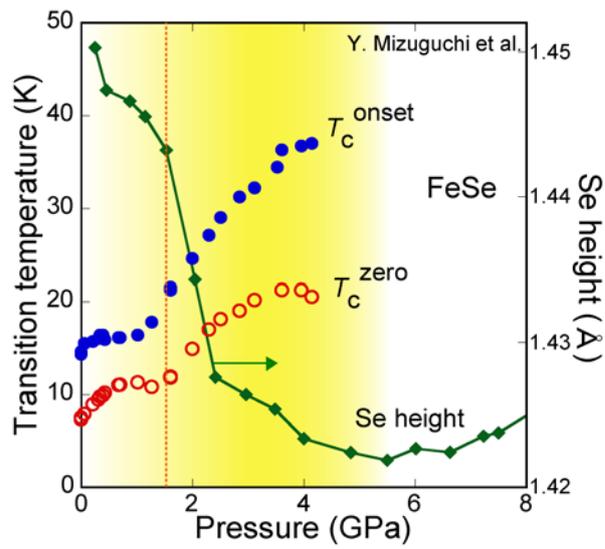

Fig. 28

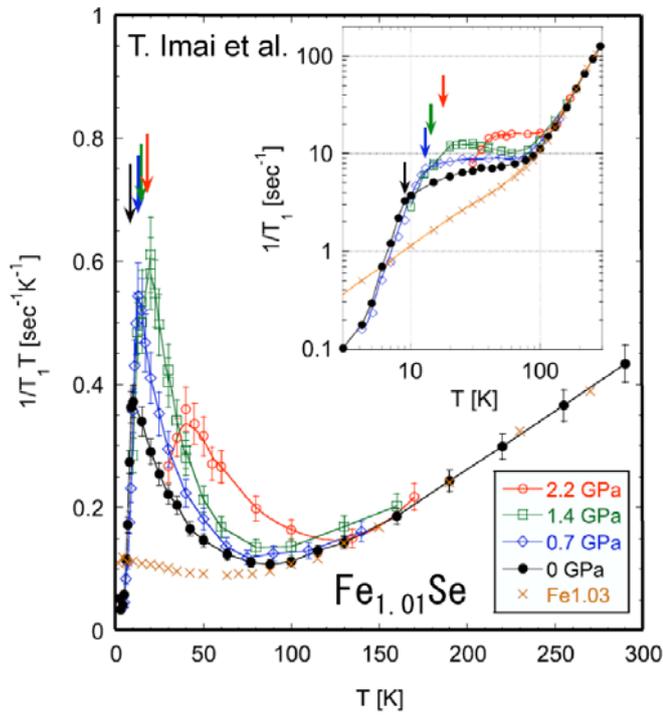

Fig. 29

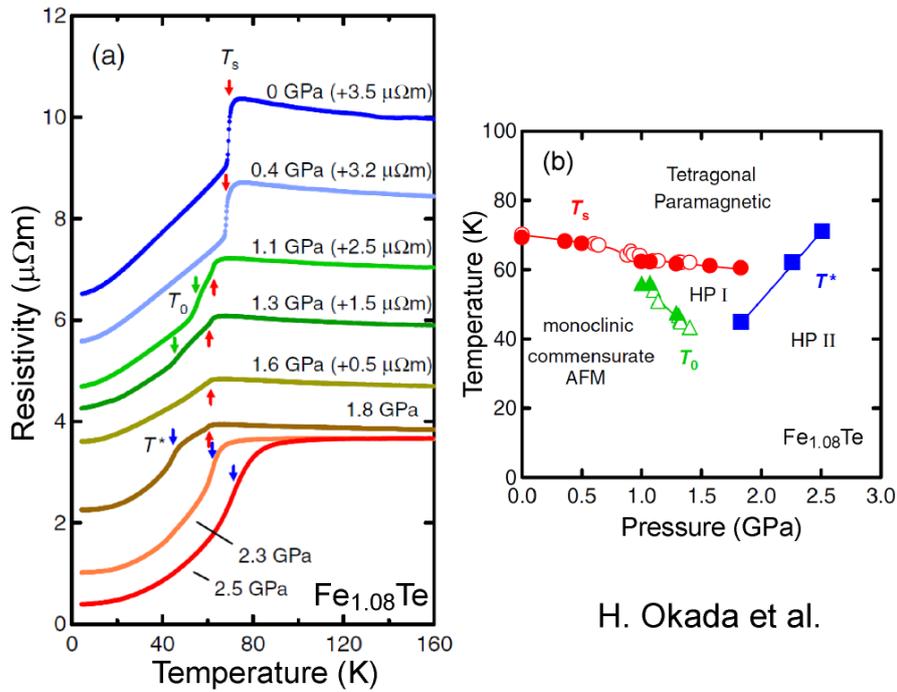

Fig. 30

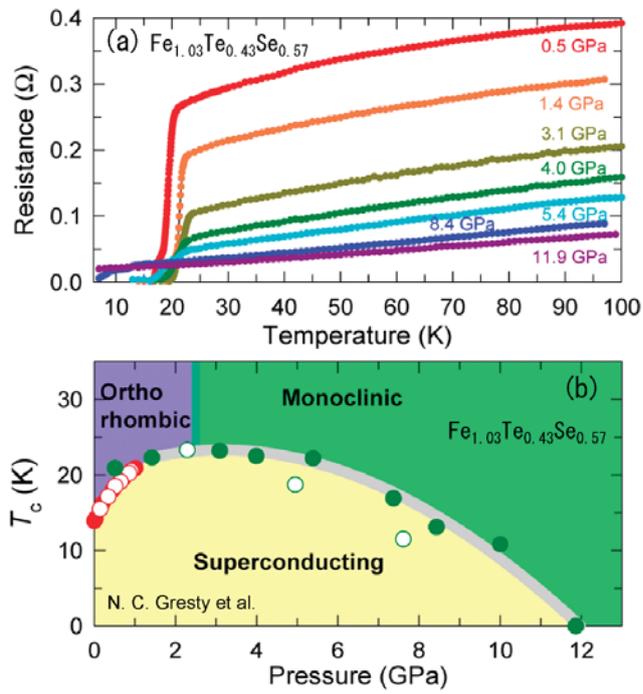

Fig. 31

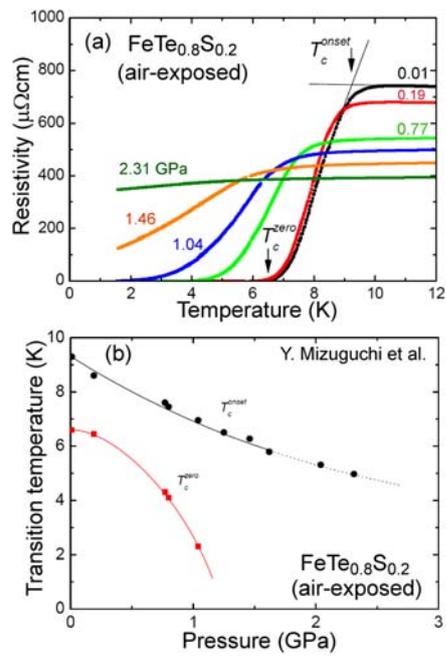

Fig. 32

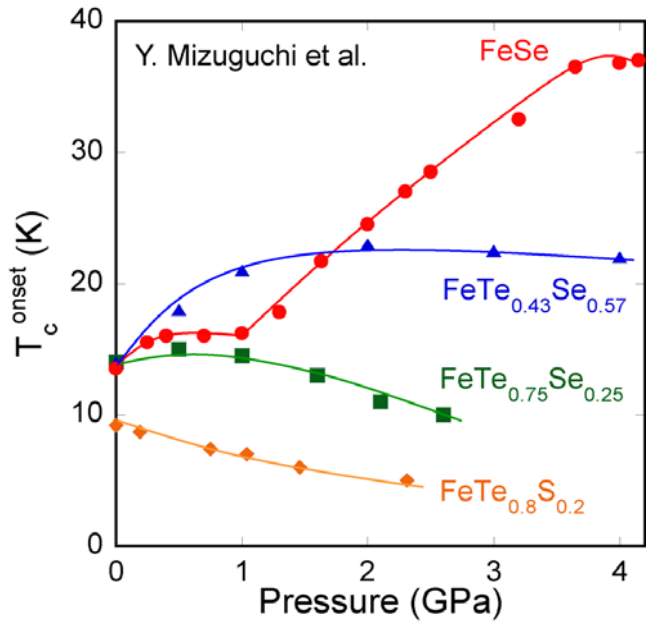

Fig. 33

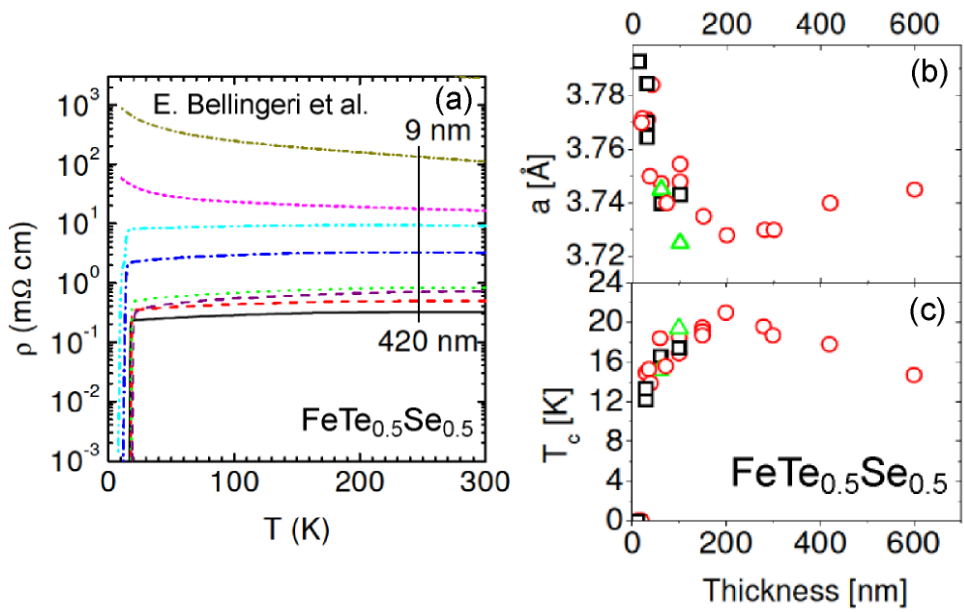

Fig. 34

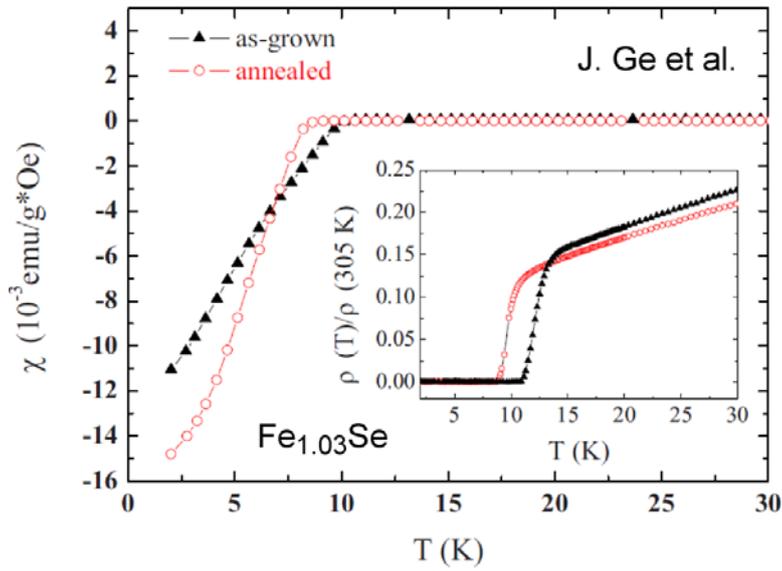

Fig. 35

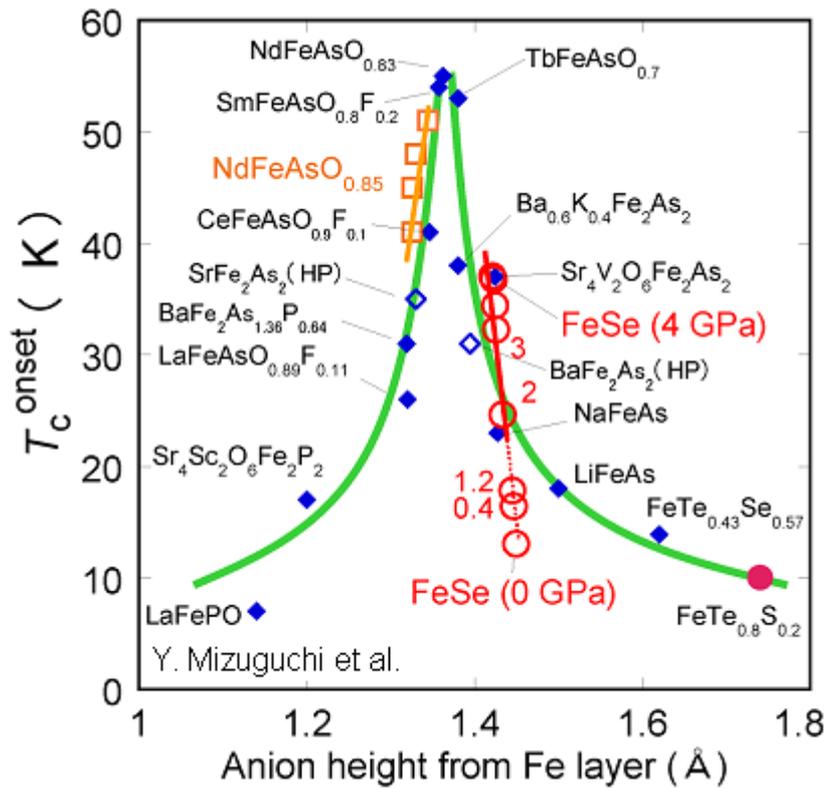

Fig. 36

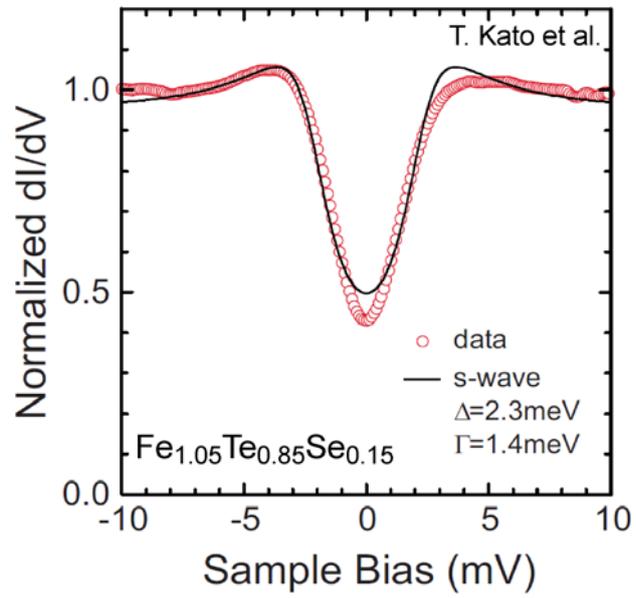

Fig. 37

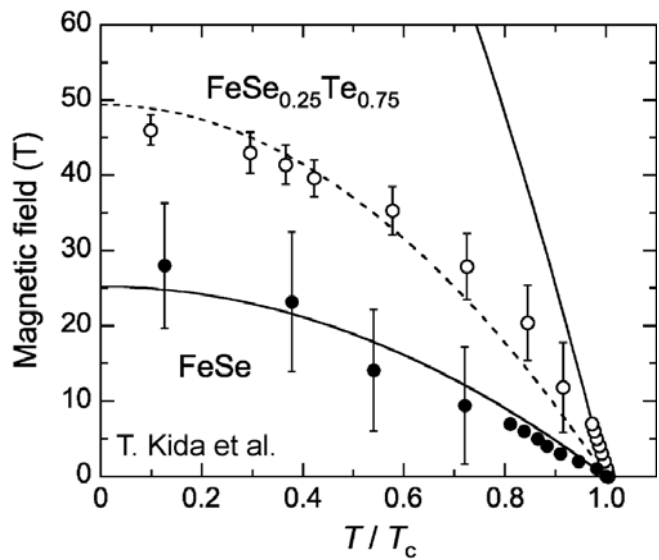

Fig. 38

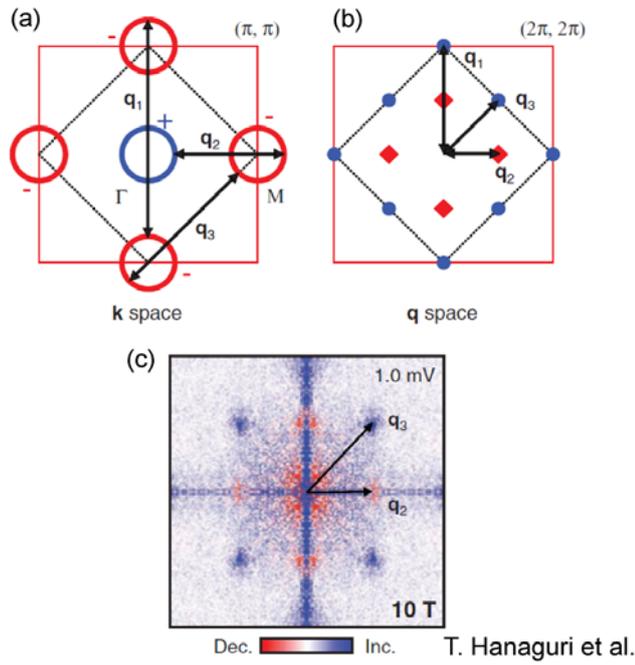

Fig. 39

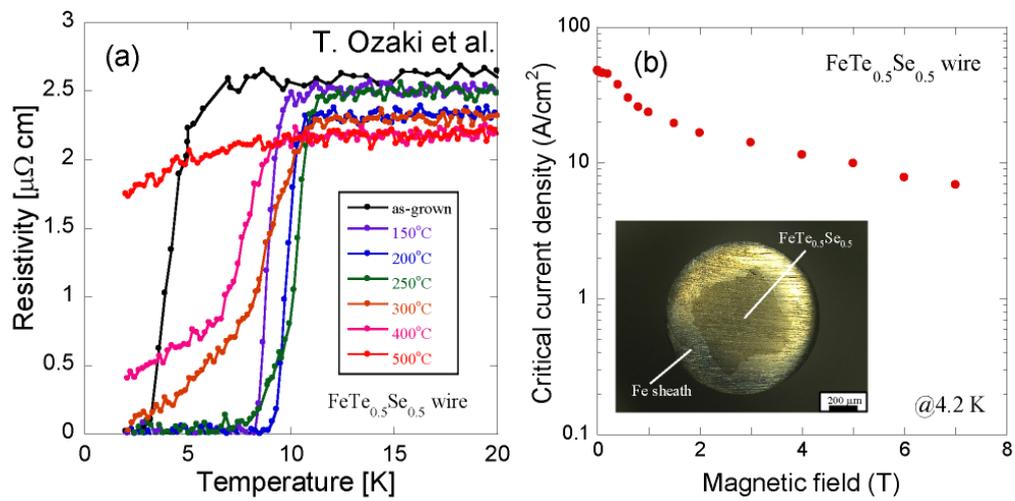

Fig. 40

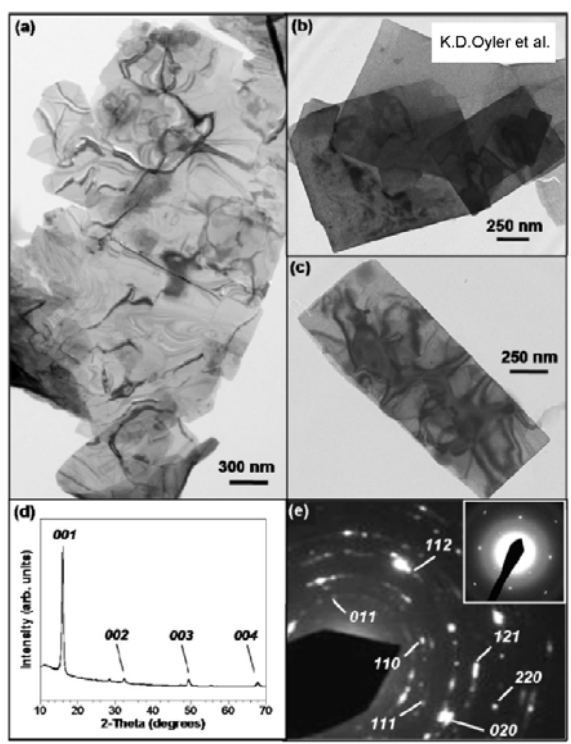

Fig. 41

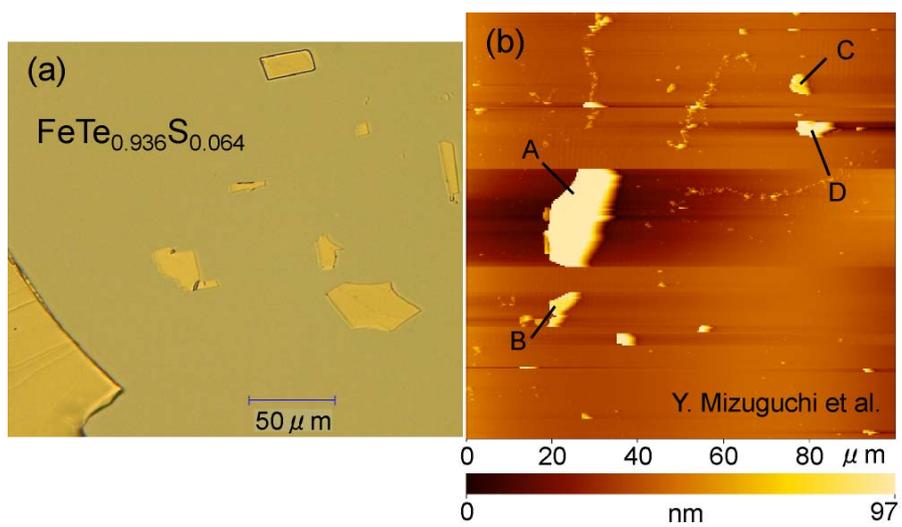